\documentclass[jcp,aip,showpacs,amsmath,amssymb,unsortedaddress,12p]{revtex4-1}

\usepackage{graphicx}

\begin{document}

\preprint{PREPRINT}

\title{Constant-force approach to discontinuous potentials}

\author{Pedro Orea}
\affiliation{Programa de Ingenier\'{\i}a Molecular, Instituto Mexicano del Petr\'{o}leo, Eje Central L\'{a}zaro C\'{a}rdenas 152, 07730 M\'{e}xico D.F., M\'exico.}

\author{Gerardo Odriozola} 
\email{godriozo@imp.mx (corresponding author)}
\affiliation{Programa de Ingenier\'{\i}a Molecular, Instituto Mexicano del Petr\'{o}leo, Eje Central L\'{a}zaro C\'{a}rdenas 152, 07730 M\'{e}xico D.F., M\'exico.}

\date{\today}

\begin{abstract}
Aiming to approach the thermodynamical properties of hard-core systems by standard molecular dynamics simulation, we propose setting a repulsive constant-force for overlapping particles. That is, the discontinuity of the pair potential is replaced by a linear function with a large negative slope. Hence, the core-core repulsion, usually modeled with a power function of distance, yields a large force as soon as the cores slightly overlap. This leads to a quasi-hardcore behavior. The idea is tested for a triangle potential of short range. The results obtained by replica exchange molecular dynamics for several repulsive forces are contrasted with the ones obtained for the discontinuous potential and by means of replica exchange Monte Carlo. We found remarkable agreements for the vapor-liquid coexistence densities as well as for the surface tension. 
\end{abstract}


\maketitle

\section{Introduction}

From the very beginning of molecular simulation, discontinuous pair potentials have played an important role~\cite{Alder55,Alder57,Wood57}. Examples of popular systems containing a hard-core interaction are, among others, hard spheres~\cite{Lebowitz59,Carnahan69,Mulero08}, square well~\cite{Henderson67,Rotendber65,Jackson92,Rio02}, attractive Yukawa ~\cite{Henderson78,Frenkel94,Duda07,Benavides12}, and Sutherland fluids~\cite{Largo00,Camp01,Nezbeda10}. These systems are frequently found in theoretical studies~\cite{Lebowitz59,Carnahan69,Henderson67,Henderson78,Largo00} and, in the frame of simulations, are mostly accessed by means of Monte Carlo (MC) techniques~\cite{Wood57,Jackson92,Rio02,Frenkel94,Camp01}. Naturally, molecular dynamics (MD) implementations of this kind of potentials, albeit possible, are not so frequent, since a special treatment is needed~\cite{Alder59,Chapela84,Singh03}. Moreover, regardless of the employed simulation technique, 
thermodynamical properties accessed though the virial route also require an alternative implementation to deal with discontinuities~\cite{Alejandre99,Orea03,Jackson05}. In this work we test approaching the discontinuity by a linear function with a large negative slope, i. e. a constant repulsive force acting on overlapped particles, for a twofold purpose: to approach hard potentials with standard molecular dynamics and to gain access through the virial route to the thermodynamic properties of hard systems. As we will show, by setting a sufficiently large slope we obtain a remarkable agreement with the hard-core system of reference while avoiding special treatments. 

The surface tension is a very sensitive quantity which is frequently accessed through the virial route (for other routes see references~\cite{Jackson05,Bryk07,Kofke07,Miguel08}). It strongly depends on slight changes of the pair-potential, as well as on changes of system size and liquid-vapor surface area when they are not large enough~\cite{Orea05,Malfreyt09,Janecek09}. Thus, for a given system, when surface tension and coexistence data of independent sources match each other, one can expect a match for other thermodynamic properties. Furthermore, obtaining the surface tension turns out to be more computationally demanding as the potential range is shortened~\cite{Rendon06,Odriozola11,Minerva12}. This is specially true when dealing with hard-cores. On the one hand, a shortening of the potential range leads to a decrease of the critical temperature making sampling more difficult. On the other hand, it gives rise to a sharper definition of the main peak of the radial distribution function for the liquid phase.
 Whenever 
the employed methodology fails to correctly capture the contact density for the liquid phase, it will certainly introduce deviations in the coexistence and surface properties.      

In view of the above mentioned points, we are testing the constant-force approach to the discontinuous part of a triangle potential~\cite{Rocco64,Rocco65,Betan07,Betan08,Haro12,Koyuncu11} of short range. We are using a triangle potential since, in contrast with the square well potential, it has only one discontinuity. Furthermore, it does not introduce truncation issues, contrasting with Yukawa and Sutherland. The potential is given by   
\begin{equation}
\label{triangle}
u(x)=\left\{\begin{array}{ll}\infty, & \mbox{ for $x \leq 1,$} \\
             \epsilon(x-\lambda)/(\lambda-1) , & \mbox{ for $ 1 < x \leq \lambda, $}\\
               0,     & \mbox{ for $\lambda<x,$}
             \end{array} \right. 
\end{equation}
where $\epsilon$ and $\lambda$ are its depth and range, respectively. In addition, $x=r/\sigma$ is the reduced distance where $\sigma$ is the hard-core diameter. We are setting $\lambda=1.5$. The proposed potential to approach equation (\ref{triangle}) is
\begin{equation}
\label{doubletriangle}
u(x)=\left\{\begin{array}{ll} \epsilon \alpha(1-x) - \epsilon , & \mbox{ for $x \leq 1,$} \\
             \epsilon(x-\lambda)/(\lambda-1) , & \mbox{ for $ 1 < x \leq \lambda, $}\\
               0,     & \mbox{ for $\lambda<x,$}
             \end{array} \right. 
\end{equation}
where the hard-core contribution to the pair potential was substituted by a linear function with slope $du(r)/dr=- \epsilon \alpha / \sigma$, with $\alpha>0$. Hence, the force for $r\leq \sigma$ is $\epsilon \alpha /\sigma$, repulsive, and constant. The potential has a double-triangle shape, it is continuous, and it presents a discontinuity in its first derivative at $x=1$. Thus, it avoids a zero-force point at its minimum and a weak-force region for small overlaps. Moreover, the hard-core is recovered for $\alpha \rightarrow \infty$. So, producing results for increasing $\alpha$ values and extrapolating them towards $\alpha^{-1} \rightarrow 0$ can be seen as a strategy to approach the hard-core limit. In view of the obtained results this extrapolation is not even necessary. 

It is worth mentioning that the constant-force approximation contrasts previous works where the core hardness is modeled by adding a high-exponent power-law contribution~\cite{Minerva04,Minerva12,Jover12}. This power-law is usually implemented for all $x$, so that the potential is infinitely differentiable (smooth) at $x=1$. Consequently, this approximation produces a zero-force point at the potential minimum and small repulsive forces for small overlaps. Finally, it should be noted that the proposed constant-force repulsion can be easily adapted to other hard potentials. For instance, for hard spheres the approximation turns into $u(x)= \epsilon\alpha(1-x)$ for $x\leq 1$ and $0$ for $x > 1$.

\section{Simulation details}

Following previous work~\cite{Odriozola11}, we combine the replica exchange Monte Carlo (REMC) method~\cite{Lyubart92,Marinari92,Hukushima96} with the slab technique~\cite{Chapela77} to produce the reference data (using equation (\ref{triangle})). They consist of vapor-liquid coexistence densities, as well as the corresponding surface tension values. To the best of our knowledge, there are no surface tension values reported in the literature for the triangle potential (this potential has not been extensively studied~\cite{Haro12}). Replica exchange is employed to improve sampling at low temperatures as usual. Thus, $M$ replicas of the system are simultaneously considered, being each one at a different temperature and allowing for swap moves. The technique enhances the sampling of the low temperature coexistence regions. Swaps are based on the definition of an extended ensemble, $Q_{ext}=\prod_{i=1}^{M} Q_{N V T_i}$, where $Q_{NVT_i}$ is the partition function of the canonical ensemble of the system at 
temperature $T_i$, volume $V$, with $N$ particles. The number of replicas, $M=19$, equals the number of different temperatures defining the extended ensemble. To fulfill detailed balance, the acceptance probability for swap trials (performed between adjacent replicas only) is $P_{acc}\!=\!min(1,\exp[(\beta_j- \beta_i)(U_i-U_j)])$, where $U_i-U_j$ is the potential energy difference between replicas $i$ and $j$, and $\beta_i-\beta_j$ is the difference between the reciprocal temperatures $i$ and $j$. Adjacent temperatures should be close enough to provide large swap acceptance rates. 

We employ parallelepiped boxes with sides $L_x=L_y=10\sigma$ and $L_z=4 L_x$ for simulating systems containing a couple of vapor-liquid interfaces. With these parameters finite size effects are avoided~\cite{Orea05,Malfreyt09,Janecek09}. Each cell is initially set with all particles ($N=1500$) randomly placed within the liquid slab and surrounded by vacuum. The center of mass is placed at the box center. Periodic boundary conditions are set in the three directions. Particles are moved by using the Metropolis algorithm and Verlet-lists are employed to speed up calculations~\cite{Verlet67}. Simulations are carried out in the vapor-liquid region, so the highest temperature is set close to and below the critical temperature. Other temperatures are fixed by following a geometrically decreasing trend. The replicas are equilibrated by $10^{7}$ MC steps. The thermodynamic properties are calculated by considering additional $4\times10^{7}$ MC steps, to produce the data for the hard-core triangle potential. All 
results are given in dimensionless units: $\rho^*=\rho \sigma^3$, $T^*=k_BT/\epsilon$, $\gamma^*=\gamma \sigma^2/\epsilon$, where $\rho$ is the number density, $k_B$ is the Boltzmann constant, and $\gamma$ is the surface tension.     

The molecular dynamics data for potential~(\ref{doubletriangle}) are obtained by using Gromacs~\cite{Spoel95,Spoel01,Spoel05,Spoel08}. This package has the enormous advantages of being free, flexible, and fast. As an example of its flexibility, we are implementing equation (\ref{doubletriangle}) as an external input table (the code was not modified). We are also making use of replica exchange, in this case replica exchange molecular dynamics (REMD), by setting certain external flags. We are setting the same conditions as for the REMC reference except for: $\alpha=100$, 200, 400, and 800; a time step of 0.0002 in reduced units for all cases; a V-rescale thermostat to keep a constant temperature for each ensemble~\cite{Bussi07}; a number of replicas $M=21$; and $4\times10^{7}$ MD cycles. Another slight difference is that our REMC implementation shifts the set temperatures during equilibration to achieve an approximately constant swap rate, while they are kept fixed with REMD. 

It should be noted that we use the same time step for all $\alpha$. In general, one should set a decreasing time step for increasing $\alpha$. We do this to show that the employed time step is sufficiently small to yield consistent thermodynamic results. For $\alpha=100$ the repulsive force is 50 times larger than the attractive one, allowing for a larger time step than 0.0002. On the other hand, the set time step for the largest $\alpha$ values may lead to deviations of dynamical properties. To be safe, another calculation with a time step of 0.0001 was also performed for $\alpha=800$ to detect possible deviations of thermodynamic properties. We detect no statistically significant differences when decreasing the time step.

\begin{figure}
\resizebox{0.5\textwidth}{!}{\includegraphics{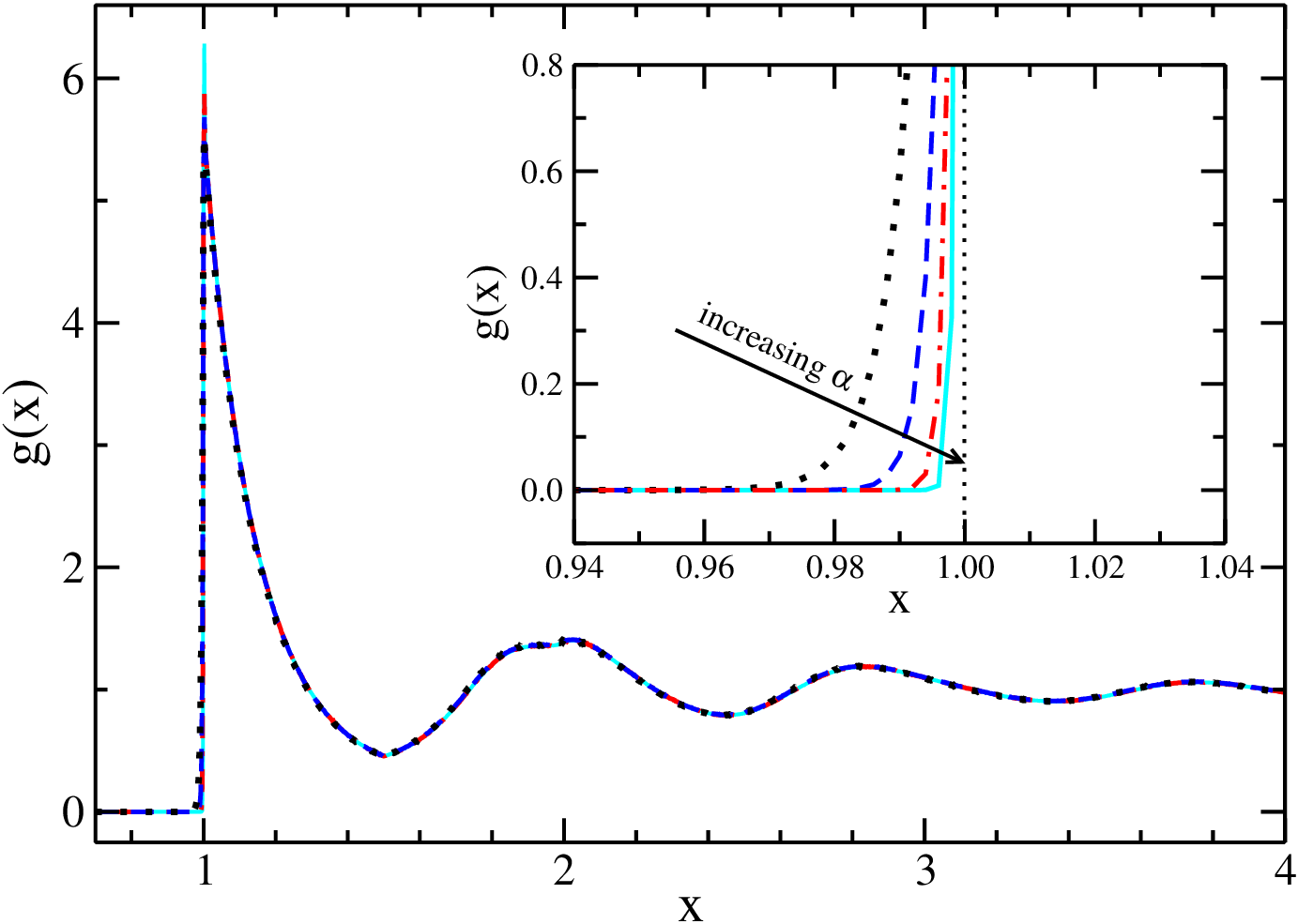}}
\caption{\label{gdrs} Radial distribution functions for equation (\ref{doubletriangle}) and the smallest temperature, as obtained by REMD. Black dotted, blue dashed, red dashed-dotted, and cyan solid lines correspond to $\alpha=100$, 200, 400, and 800, respectively. The inset zooms in on the same data. Inside it, the arrow points to the $\alpha$ increasing direction and the dotted vertical line represents the hard-core limit.}
\end{figure}

For both, REMC and REMD, the surface tension is obtained from the difference of the ensemble averages between the normal and tangential components of the pressure, i. e., 
\begin{equation}
\label{virial}
\gamma =\frac{L_z}2\Bigg\{\big<P_{zz}\big> -
\frac 12\big[\big<P_{xx}\big> + \big<P_{yy}\big>\big]\Bigg\},
\end{equation}
where $P_{ii}$ are the diagonal components of the pressure tensor. The factor $1/2$ is due to the existence of two interfaces in the system. The implementation of equation (\ref{virial}) is generally known as the virial route. For discontinuous potentials, the pressure components are obtained by following the methodology given in previous works~\cite{Alejandre99,Orea03,Odriozola11}. This well-proven, but somewhat tedious approximation involves an extrapolation procedure. This methodology is avoided by implementing equation (\ref{doubletriangle}). Note that using equation (\ref{doubletriangle}) is independent of the simulation method (MC can be used).

\section{Results and discussion}

Before analyzing the coexistence densities and surface tension data, it is instructive to focus on the radial distribution functions obtained for $\alpha=100$, 200, 400, and 800, as shown in figure~\ref{gdrs}. It can be seen there are practically no differences among them. Hence, thermodynamic properties from the different approximations are expected to be similar. The only difference is observed for the main peak, which slightly sharpens and increases its height with increasing $\alpha$. This issue is related to the occurrence of overlapping configurations, whose frequency and range decrease as $\alpha$ rises. The inset of the radial distribution function close to contact clearly shows this trend. We observed that $g(0.99)=$ 0.6, 0.06, 8$\times 10^{-4}$, and 2$\times 10^{-7}$ for $\alpha=100$, 200, 400, and 800, respectively. The frequency and range of the overlaps are small in all cases.

\begin{figure}
\resizebox{0.5\textwidth}{!}{\includegraphics{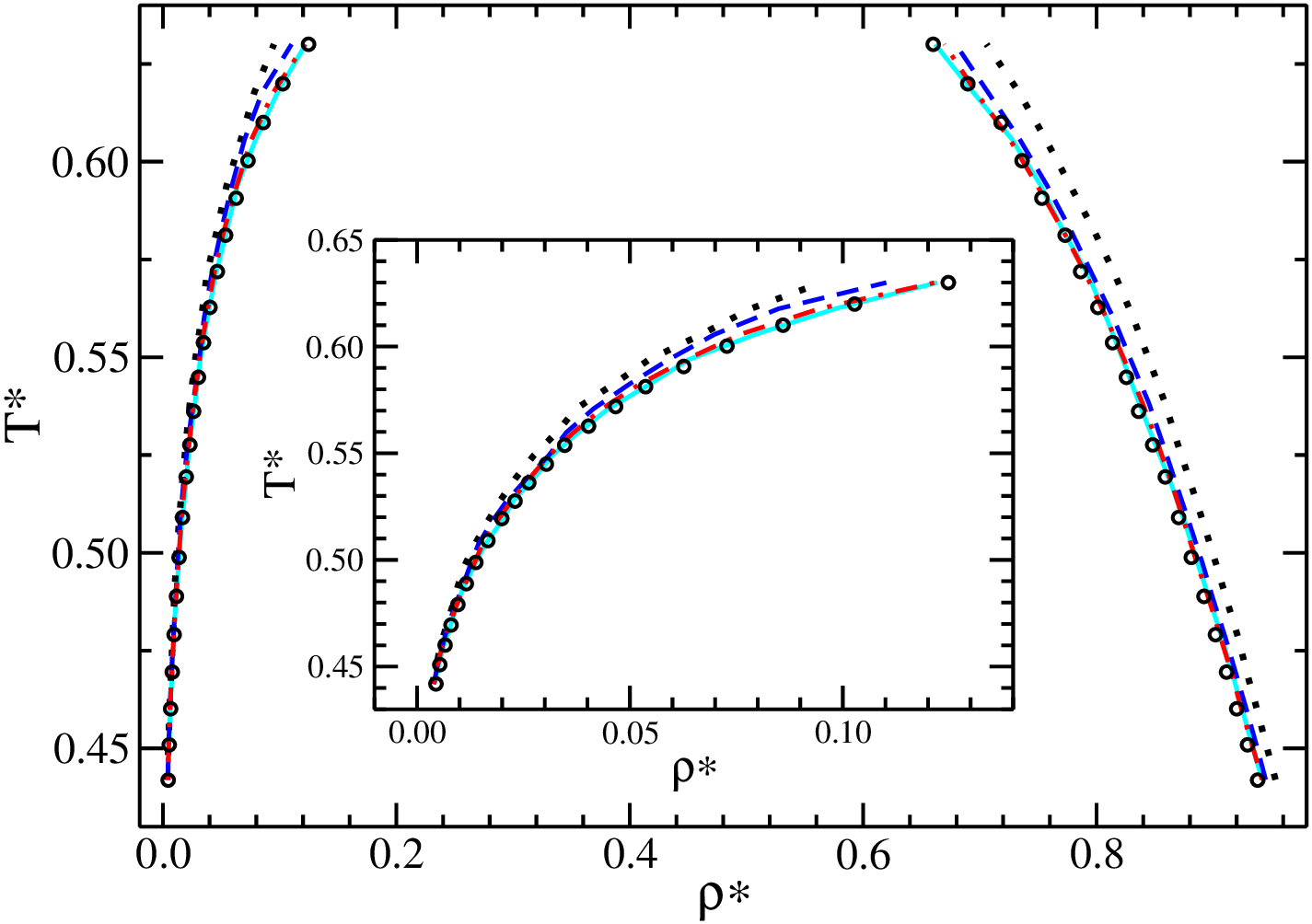}}
\caption{\label{coexistence} Coexistence densities for the hard-core triangle potential (black circles) as obtained by REMC, together with REMD data for equation~(\ref{doubletriangle}) with $\alpha=100$ (black dotted lines), 200 (blue dashed lines), 400 (red dashed-dotted lines), and 800 (cyan solid lines). The inset zooms in on the vapor branch.}
\end{figure}

The coexistence densities are obtained directly from appropriate averages in the different regions of the density profiles, which allows to obtain precise values for the liquid and vapor phases. The obtained results from both, the target potential (equation (\ref{triangle})) and the approximation (equation (\ref{doubletriangle})) with $\alpha=100$, 200, 400, and 800 are shown in Fig.~\ref{coexistence}. As previously explained, the data for the target potential are obtained by REMC, whereas the data corresponding to equation (\ref{doubletriangle}) are obtained by REMD. Data from Betancourt et al.~well agree with REMC results and are not shown to gain clarity~\cite{Betan07}. The first thing to note is that, even for the lowest $\alpha$ considered, the target potential and its approximation yield similar results. The second point to highlight is the fact that, for $\alpha=800$, we do not detect differences with the hard-core case, accounting for the statistical error (less than 4$\%$ for all data, although 
considerably smaller at low temperatures). This contrasts with the modeling of the core hardness as a power-law, $(1/x)^n$. In this case, even considering extremely large exponents, $n>200$, the differences in the coexistence properties between the approximation and the hard-core case do not vanish~\cite{Minerva12}. A final remarkable issue is that the data series for $\alpha=400$ is indistinguishable from the one for $\alpha=800$. Thus, an extrapolation of the results for $\alpha^{-1}\rightarrow 0$ is not necessary.

\begin{figure}
\resizebox{0.5\textwidth}{!}{\includegraphics{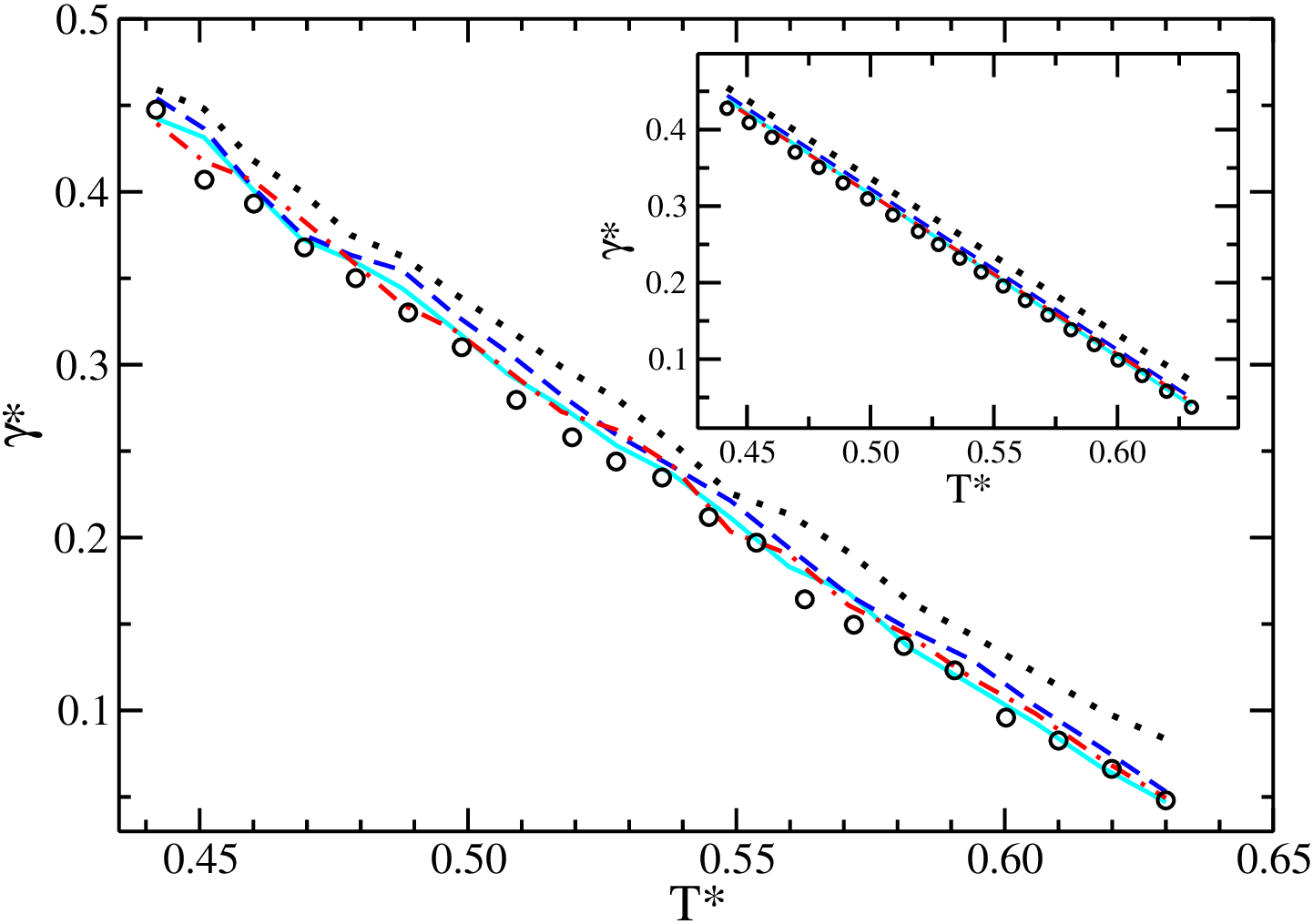}}
\caption{\label{surfacetension} Surface tension curves for the hard-core triangle potential as obtained by REMC, as well as REMD data for potential~(\ref{doubletriangle}) with $\alpha=100$, 200, 400, and 800. Symbols are in correspondence with Figure~\ref{coexistence}. The inset shows fitting lines to the simulation data sets to avoid noise and make the differences among them clearer. }
\end{figure}

The surface tension data, corresponding to the liquid-vapor coexistence curves shown in Fig.~\ref{coexistence}, are given in Fig.~\ref{surfacetension}. Symbols are also in correspondence with those used in Fig.~\ref{coexistence}. The inset shows the linear fits to the data sets, to avoid noise and make the comparison easier. It is worth mentioning that the maximum deviation between a particular point and the fitted line is $0.016$ for all cases. This is a tiny absolute deviation. However, the relative deviation is not so small due to the very small surface tension intrinsic to short range potentials. As for the vapor and liquid density branches, all data sets are very close to each other. Again, noticeable differences appear between the $\alpha=100$ case and the other cases. These data are shifted an approximately constant quantity towards larger values, $\simeq 0.02$ according to the average difference between the linear fits to both series (see the inset of Fig.~\ref{surfacetension}). Thus, relative 
differences increase for increasing temperature, being considerably large close to the critical temperature. There is also a probable statistically significant difference between data from $\alpha=200$ and $\alpha=400$. In this case the shift would be $\lesssim 0.005$. We detect no differences when further increasing $\alpha$. Again, an extrapolation towards $\alpha^{-1} \rightarrow 0$ is, in our view, not justified. Finally, a very good agreement is obtained when comparing the results from the target potential with those from the approximation with $\alpha \geq 400$. 

\section{Conclusions}

We have compared results from the constant-force approach to those of the hard-core triangle potential finding an excellent agreement for $\alpha \geq 400$. This was obtained by setting the repulsive hard-core forces over two-hundred times the attractive ones, so that $g(0.99) \lesssim 0.001$. That is, most overlaps are very small. This guarantees, at least in the studied cases, an excellent agreement between the results from the approximation and the hard-core limit. We expect this rule of thumb (setting the repulsive force so that $g(0.99) \lesssim 0.001$ to yield hard-core limit properties) to hold for all hard-core potentials and thermodynamic conditions. With this simple idea one can perform MD simulations of model systems with discontinuous potentials without the need of employing special treatments, even to evaluate virial dependent properties. Actually, user friendly and flexible packages such as Gromacs can be directly used to study discontinuous potentials. Finally, we should also mention that the 
discontinuity generated 
by the necessary truncation of continuous potentials at the cutoff distance, which introduces differences between MD and MC results, can also be handled by following a similar idea.         

\begin{acknowledgments}
The authors thank The Molecular Engineering Program of IMP. GO acknowledges CONACyT of M\'exico for financial support though grant No.~169125.
\end{acknowledgments}


\end{document}